\documentclass[12pt]{article}

\DeclareUnicodeCharacter{02DA}{\degree}

\usepackage{booktabs}
\usepackage{graphicx} 
\usepackage[letterpaper,top=2cm,bottom=2cm,left=2cm,right=2cm,marginparwidth=1.75cm]{geometry}
\usepackage{enumitem}

\usepackage{xcolor}
\usepackage{amsmath}
\usepackage{authblk} 
\usepackage{gensymb}
\usepackage{siunitx}
\usepackage{soul}
\usepackage{multirow}
\usepackage{gensymb}

\usepackage{url}

\usepackage{lineno}

\usepackage{tikz}
\usetikzlibrary{arrows.meta, positioning}
\usetikzlibrary{calc,fit}

\newcommand{\inpt}[1]{{Input}}
\newcommand{\blck}[1]{{Block$_{#1}$}}

\tikzset{
  inputbox/.style={
    draw,
    minimum width=1cm,
    minimum height=1cm,
    align=center,
    fill=blue!10,
    },
   block/.style={
    draw,
    minimum width=1cm,
    minimum height=1cm,
    align=center,
    fill=red!30,
    },
   blockd/.style={
    draw,
    minimum width=1cm,
    minimum height=1cm,
    align=center,
    fill=orange!50,
    },
  arrow/.style={
    -{Stealth},
    thick
  },
  dotsnode/.style={
    inner sep=2pt,
    minimum size=0pt,
  },
  output/.style={
  	circle, 
  	draw, 
  	minimum size=1cm,
  	fill=green!10,
  },
}

\title{Extensive Air Showers Parameters Estimation Using Machine Learning Techniques with Simulations of the FAST Telescope}

\date{\today}
\author[1]{Jiří~Kvita\thanks{Corresponding author: \texttt{jiri.kvita@upol.cz}}}
\author[3]{Monika~Machalová}
\author[1,2]{Radek~Přívara}
\author[3]{Rostislav~Vodák\thanks{Corresponding author: \texttt{rostislav.vodak@upol.cz}}}
\author[3]{Jan~Tomeček}

\affil[1]{Palacký University in Olomouc, Faculty of Science, Joint Laboratory of Optics of Palacký University and Institute of Physics AS CR, 17. listopadu 12, Olomouc, 77900, Czech Republic}
\affil[2]{Institute of Physics of the Academy of Sciences of the Czech Republic, Joint Laboratory of Optics of Palacký University and Institute of Physics AS CR, 17. listopadu 50a, Olomouc, 77900, Czech Republic}
\affil[3]{Palacký University in Olomouc, Faculty of Science, Department of Mathematical Analysis and Applications of Mathematics, 17. listopadu 12, Olomouc, 77900, Czech Republic}

\begin{document}

\providecommand{\Conex}{\textsl{Conex}}
\providecommand{\logE}{\ensuremath{\log_{10} E / \mathrm{eV}}}
\providecommand{\Xmax}{\ensuremath{X_\mathrm{max}}}
\providecommand{\gcm}[1]{\ensuremath{#1\,\mathrm{g}/\mathrm{cm}^{2}}}

\maketitle

%\linenumbers

\abstract{We present the capabilities of the Fluorescence detector Array of Single-pixel Telescopes (FAST) observatory in the single telescope configuration as an important step towards the possible future large-field observatory for detecting ultra-high-energy cosmic rays. Reconstruction of main shower physics parameters are explored on noise-free simulated events using machine learning techniques in the challenging domain of a low-intensity transient signal. We find a very good correlation between the true and reconstructed energy of the shower even with the information from just the four photomultipliers of the single FAST telescope, and a reduced performance for the maximum of the shower development \Xmax{}, using various architectures of artificial deep and convolutional neural networks, with a comparison to a benchmark gradient boost regression model.
The resolution in the energy is found at the sub-percent level, while in \Xmax{} it is~$5\%$.
The relative difference between predicted and true values is under one percent for energy, while for \Xmax{} it ranges from $-8\%$ to $+17\%$, which can be attributed to the limited information from the single FAST telescope configuration.
The results constitute an important capabilities verification and a lesson learned with implications for established FAST prototypes as well as for more complex configurations of the FAST observatory under construction.
}

\section{Introduction}

Ever since the discovery of cosmic rays~\cite{Hess:1912srp} and the first observation of TeV-range energies~\cite{Hillas:1985is}, experiments have been built to achieve larger fields of view and ultimately gather growing numbers of highly energetic particles hitting the Earth's atmosphere, ranging from primary gamma rays and electrons to protons and heavier nuclei. The two main ground-based currently operational experiments in the domain of the observation of extensive air showers (EAS) induced by the cosmic rays are the Pierre Auger Observatory (Auger)~\cite{bib:auger} and the Telescope Array experiment (TA)~\cite{bib:TA}. 

The FAST telescope is a next-generation concept to cover a large area with relatively simple and affordable telescopes at a large spacing in order to observe rare ultra-high-energy cosmic rays with energies exceeding $10^{19}\,$eV~\cite{bib:fast}.
As a single FAST telescope contains only four large photomultipliers (PMTs) forming a four-pixel camera, the missing angular granularity must be replaced by sufficient granularity in the time structure of the detected signal, with noise levels under control.

In this work, we explore the possibilities to reconstruct a subset of the EAS parameters for a single FAST telescope using Monte Carlo simulations without noise. 
Specifically, electronic noise and night-sky background are thus neglected in the subsequent analysis.
This approach is motivated by studying the machine learning (ML) based reconstruction limits as well as the limits of a single FAST telescope, a basic unit of the future array \cite{bib:fast}. Algorithms which mitigate the noise at the trigger level are described in~\cite{KMEC2026110063}, while the sensitivity to the noise is left to be discussed within the planned FAST mini-array configuration \cite{Hamal2024}.

The challenge of reconstructing shower parameters using a single FAST telescope comprising only four PMTs is that several shower parameters such as the energy of primary particles, incident geometry, and the maximum of the shower development \Xmax{} can produce similar signals in the PMTs.
We thus restrict the prediction to the two primary physics parameters: the primary particle energy, expressed as \logE{}, and \Xmax{}.

\section{Motivation}
The reconstruction of the EAS parameters is a challenging task in many aspects. Traditional top-down reconstruction methods applicable for FAST are based on comparing the reconstructed signal with those coming from simulations with various shower parameters using, \emph{e.g.} a likelihood function \cite{Malacari:2019uqw}. The practical difficulty with this approach is the need for either a very fine grid of simulated showers for various parameters or the necessity to generate a simulation for given shower parameters in the real time during the likelihood optimisation. This makes the procedure highly CPU- and time-demanding, and sensitive to initial parameters values.

In this work, we explore complementary ML-based approaches which are time-demanding in the training phase of the classifier but then very time-efficient in classifying the individual events. Such ML-based estimators can be used either for initial parameters reconstruction as a pre-classifier before a classical likelihood technique, or as a full reconstruction tool. In addition, the likelihood techniques are not easily scalable to the case of a larger FAST array while ML-based ones still retain the short evaluation time.

\section{The FAST telescopes concept}

\begin{figure}[h!]
    \centering
    \includegraphics[width=0.33\linewidth]{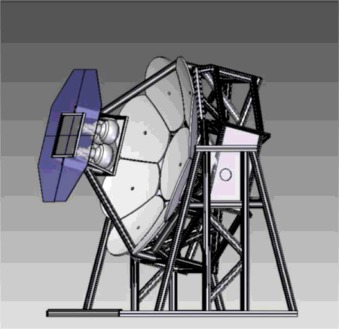} \hspace{3em}
    \includegraphics[width=0.40\linewidth]{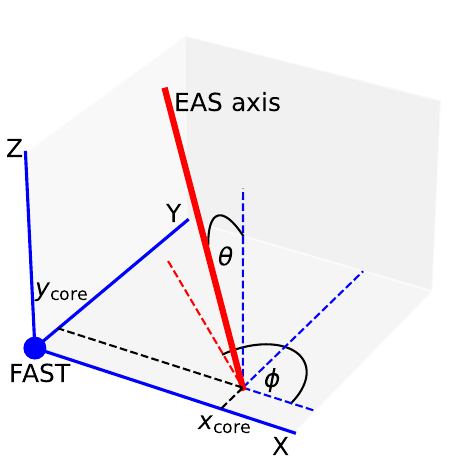}
    \caption{Left: Schematic of a single FAST telescope (without its housing), showing the support structure, the segmented mirror, the $2\times2$ pixel camera with 4 PMTs, and the UV-pass filter (blue)~\cite{bib:fast}. Right: The coordinate system used in the simulation, with the FAST telescope placed at the origin~\cite{KMEC2026110063}. 
    The geometric parameters $x_{\mathrm{core}}$, $y_{\mathrm{core}}$, azimuth angle $\phi$, and zenith angle $\theta$ are also indicated.}
    \label{fig:FAST_bw}
\end{figure}

The optical and electronics design of the FAST telescope prototypes is described, \emph{e.g.}  in~\cite{bib:fast, bib:fast_icrc2017,bib:fast_optics,bib:fast_optics_icrc2017}
with additional details in~\cite{Malacari:2019uqw}.
See the left panel of Figure~\ref{fig:FAST_bw} for a schematic of the FAST telescope design. 
In essence, the light passes through the UV-pass filter at the aperture of about $1\,\si{m}^2$ to a segmented mirror which reflects the light to a four-pixel camera slightly off the focus, in order to expose the PMTs more evenly. The total field of view is $30{\degree} \times 30{\degree}$.
More recent FAST reports can be found in \cite{Justin:2021phd} and \cite{Hamal2024, Bradfield2024, Bradfield2025}.
The FAST telescope design is being considered as a cost-effective fluorescence detector for the proposed Global Cosmic Ray Observatory \cite{GCOS:2021exh}, aiming to cover an area over $60,\!000\,\si{km}^2$~\cite{Ahlers2025}.

Similarly to fluorescence detectors of the Auger and TA observatories, the FAST telescopes also observe the fluorescence light emitted by excited molecules of atmospheric nitrogen when ionization processes from the EAS take place upon the primary particle interaction in the upper atmosphere \cite{Ave2008, Keilhauer2013}.

The telescopes are already installed and operational on both hemispheres~\cite{FAST:2023ibh}, at the TA and Auger sites, taking data steadily. Therefore, the challenge of obtaining the most accurate information from the incoming data in terms of physics shower parameters reconstruction is of an imminent importance.

\section{Simulations}

\subsection{Simulation description}

 The simulations used in this study were generated by the FAST Collaboration~\cite{Kmec2025_data,KMEC2026110063} and are used through the open data policy of the CC BY license. This open licensing grants access to the FAST Collaboration dataset, making it possible to utilize these simulations to propose a potential processing frameworks using ML methods.

In more detail, the showers are generated using the FAST framework~\cite{Malacari:2019uqw} where the FAST telescope is placed at the origin with its optical axis pointing along the $y$-axis. The generated primary particles in this study are protons.
Each shower is defined by its energy, \Xmax{}, azimuth angle $\phi$, zenith angle $\theta$, and core position [$x_{\mathrm{core}}$, $y_{\mathrm{core}}$].
The geometry of the angular and spatial parameters of the shower are shown in the right panel of Figure~\ref{fig:FAST_bw}. 
The details of the simulation are described in~\cite{Malacari:2019uqw}.

The ML techniques are applied to simulated events that reproduce realistic signal profiles, including the correlations across the PMTs. 
All input parameters are drawn from continuous uniform distributions over predefined intervals, except for \Xmax{}, for which a more realistic distribution is applied. 
The energy is generated uniformly in \logE{} between 17.5 and 20.5, the azimuth angle uniformly in the range from $-180$\degree{} to $180$\degree{}, and the zenith angle between 0\degree{} and 85\degree{}. 
Showers with zenith angles above 85\degree{} are excluded to avoid unrealistic horizontal showers. 
The core positions, $x_{\mathrm{core}}$ and $y_{\mathrm{core}}$, are generated within intervals [$-12,\!500\,\si{m}; 12,\!500\,\si{m}$] and [$0\,\si{m}; 25,\!000\,\si{m}$], respectively.
Only events with core positions located within a radius of $25,\!000\,\unit{m}$ from the FAST telescope and within its field of view of $30$\degree{}, horizontally extended by $15$\degree{} on both sides, are considered.
This extension ensures that EASs whose ground impact locations lie outside the field of view are still included in the dataset, as they may remain observable in the atmosphere above.

\begin{figure}[h!]
    \centering
    \includegraphics[width=0.40\linewidth]{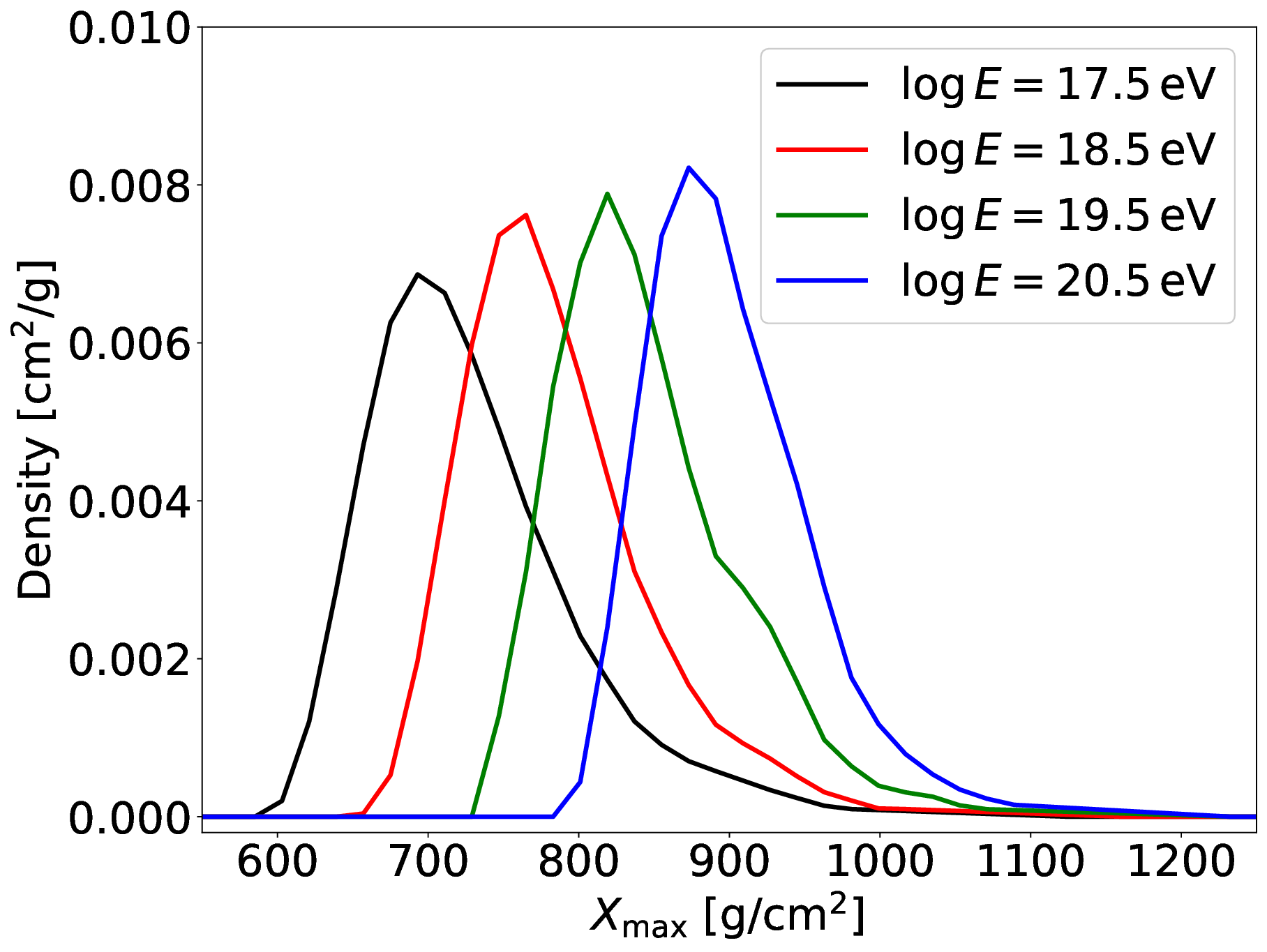}
    \caption{Probability density functions for the \Xmax{} variable as obtained from the \Conex{} simulation for an example of four fixed primary particle energies between \logE{} of 17.5 -- 20.5.}
    \label{fig:conex_Xmax_profiles}
\end{figure}

In order to simulate events with realistic physics shower profiles, for each shower energy the \Xmax{} values are drawn from a probability density function obtained from the \Conex{}~\cite{Bergmann:2006yz} generator, see
Figure~\ref{fig:conex_Xmax_profiles} for an example for several energies.
For energies between the discrete points of the \Conex{} simulation, a linear interpolation between the profiles is performed.
In total, $1,\!416,\!236$ EAS simulations were generated. All EAS simulations used in this manuscript have been also used in a different, trigger-related study in~\cite{KMEC2026110063} and are available in~\cite{Kmec2025_data}.

While the simulated \Xmax{} values follow a realistic distribution, this does not apply to the primary energy, which is sampled uniformly.
In reality, the cosmic-ray flux is highly non-linear and decreases rapidly at the highest energies.
For example, particles with energies above $10^{19}\,\si{eV}$ arrive at Earth with a rate of roughly one event per square kilometer per century.
Since our goal is to reconstruct events at highest energies, it is necessary to ensure adequate coverage of the parameter space in this area of interest.
Using a realistic energy spectrum would extremely limit the number of high-energy events in the dataset, making reliable predictions impossible.

Note that the simulations used in this analysis do not include several sources of systematic effects, namely noise coming from the electronics and the night-sky background, atmospheric non-homogeneities (\textit{e.g.} aerosols), and PMT non-uniformities.
While PMT non-uniformities can be readily implemented within the FAST framework as it can be set according to realistic telescope-specific variations, the inclusion of noise backgrounds and atmospheric non-homogeneities is less straightforward.
This is discussed in more detail in the Discussion section.

\begin{figure}[h!]
    \centering
    \includegraphics[width=0.60\linewidth]{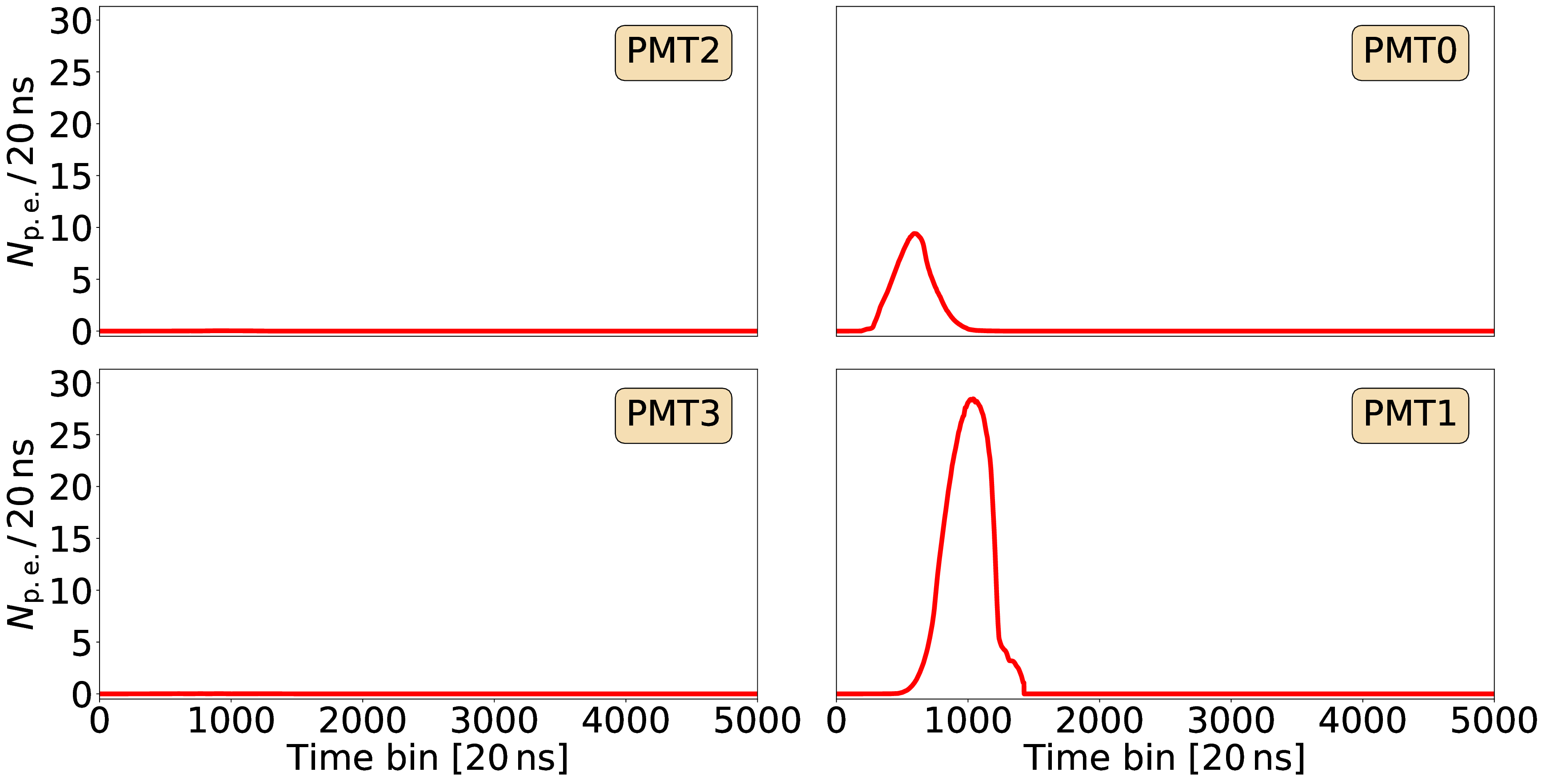}
    \caption{Example of the four FAST traces corresponding to a single simulated EAS generated with parameters 
     $\mathrm{log} E/\unit{eV} = 19.99$, $X_{\mathrm{max}} = 864.20 \, \unit{g/cm}^2$, $\phi= -125.75 \degree$, $\theta=5.92 \degree$, $x_{\mathrm{core}} = 2,\!946 \,\unit{m}$, and $y_{\mathrm{core}} = 11,\!720 \,\unit{m}$.
     The PMT layout corresponds to the sky view: the upper two PMTs observe elevation angles $15\degree-30\degree$, while the lower two observe elevation angles $0\degree - 15\degree$. 
     The signal, in units of the number of photoelectrons, is clearly visible in PMT0 and PMT1, while no signal is observed in PMT2 and PMT3.}
    \label{fig:EAS}
\end{figure}

The resulting EAS simulation consists of four individual time series, one for each PMT.
In the following, these time series are referred to as traces.
The signal in each time bin is expressed in units of the number of photoelectrons ($N_\mathrm{p.e.}$). An example of a simulated signal is shown in Figure~\ref{fig:EAS}.

\subsection{Showers filtration for a non-trivial signal}

To ensure that only showers with a~non-trivial signal are included, the following preselection criteria are applied before the machine-learning procedures.

\begin{enumerate}
    \item[C1] Maximum amplitude requirement:
    \begin{itemize}
    
        \item We require the signal maximum over all traces to be $1\,N_\mathrm{p.e.}$ $\leq$ \verb+max(traces)+ $\leq$ 2000$\,$N$_\mathrm{p.e.}$. This is motivated by the following considerations: 
        \begin{itemize}
            \item The minimum condition is due to the fact that signals with maxima of $0$ -- $1\,N_\mathrm{p.e.}$ are simply too weak for practical purposes, considering realistic observatory trigger and noise conditions \cite{KMEC2026110063}. 
    
            \item The maximum requirement is motivated by over-saturation of realistic PMTs, as such signals no longer carry useful information. 
            
            \item By applying the above minimum and maximum amplitude conditions we select $496,\!332$ events from the original $1,\!416,\!236$, \emph{i.e.} about one third.
        \end{itemize}
    
    \end{itemize}
    
    \item[C2] Removal of events with a too narrow signal in time:
    \begin{itemize}
        \item We filter out events based on the signal peak width relative to the trace signal maximum, defining a threshold of $0.1$ times the maximum. A fixed threshold is not used in order to remove narrow events which are similar in shape, but not in amplitude. The relative cut also proved to be more robust.
        
        \item We define a width of the event in each trace as 
        the largest number of consecutive time bins where the signal is above the relative threshold.
        We then select only those events in which at least one trace satisfies both following conditions: the width is greater than 10 and the maximum is greater than 1$\,N_\mathrm{p.e.}$.

        \item The narrow signal criterion removes $5,\!409$ events. 
    \end{itemize}
    
    \item[C3] Last, only samples where \Xmax{} is between \gcm{600} and \gcm{1,\!000} were considered to avoid sparsely populated regions in the parameter space and statistical fluctuations therein, which reflects the usage of a realistic \Xmax{} distribution.
    This condition was met by $477,\!141$ samples.
\end{enumerate}
The total selection efficiency based on the conditions described above is $33.7\%$, see Table~\ref{tab:filtration} for a summary.

\begin{table}[h]
\centering
\begin{tabular}{l|r|r}
\toprule
Selection criterion & removed & remaining \\
\hline
none & $-$ & $1,\!416,\!236$ \\
C1 (maximum amplitude) & $919,\!904$ & $496,\!332$ \\
C2 (narrow signals removal) & $5,\!409$ &$490,\!923$ \\
C3 (\Xmax{} cut) & $13,\!782$ &$477,\!141$ \\
\bottomrule
\end{tabular}
\caption{Number of showers after the filtration steps. In total, $477,\!141$ showers remained after applying all preselection criteria C1--C3.}
\label{tab:filtration}
\end{table}

\begin{figure}[h!]
    \centering
    \includegraphics[width=0.9\linewidth]{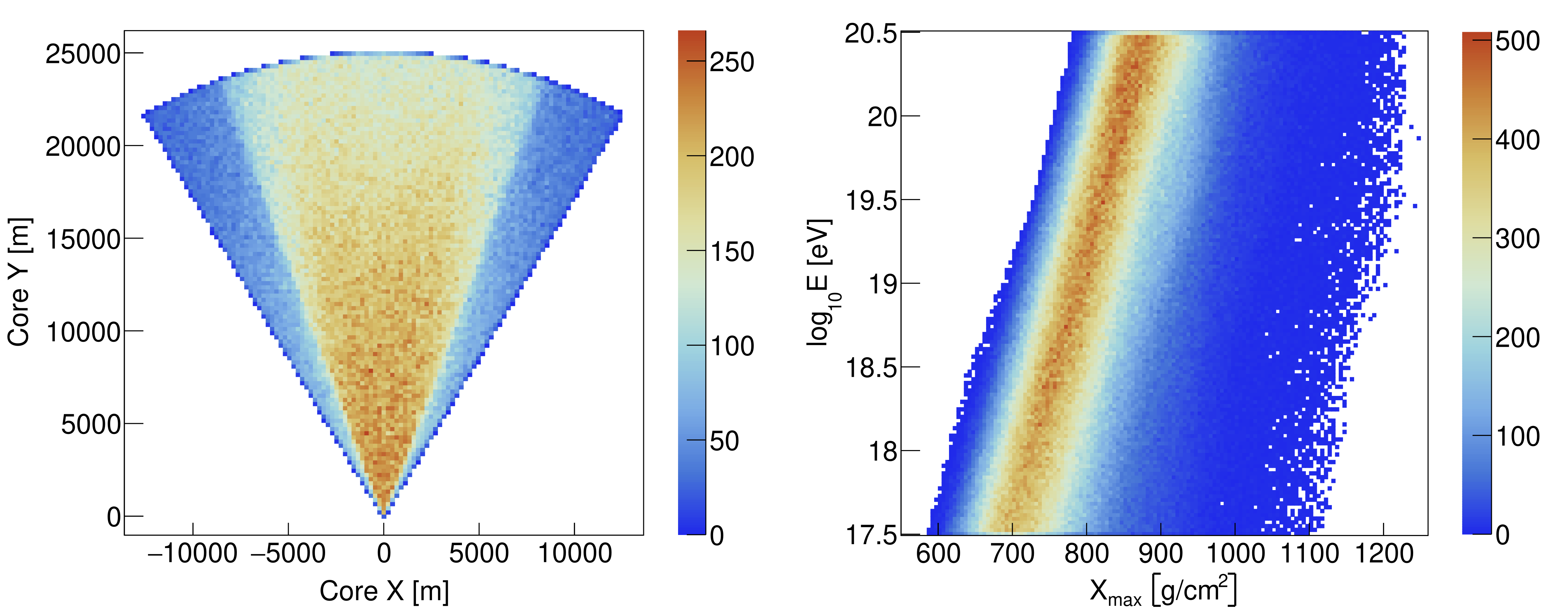}
    \caption{Overview of the generated parameter space for preselected showers used in the study: The scatter plot of the shower impact core position w.r.t. the FAST telescope location at $[0,0]$ (left) and the scatter plot of energies and \Xmax{} (right). The density of simulations in parameter space is indicated by the colour scale on the right.}
    \label{fig:gen_phasespace}
\end{figure}

The above preselection criteria lead to the shower geometrical parameters as depicted in~Figure~\ref{fig:gen_phasespace}.
It is worth noting that the core coordinates correspond to the impact point of the shower at ground level (see the right panel of Figure \ref{fig:FAST_bw}); this is why the FAST telescope can observe showers whose ground impact points lie outside the nominal geometric field of view of $30{\degree} \times 30{\degree}$.

\section{EAS Parameters reconstruction using machine learning techniques}
\subsection{Dataset description}
The entire simulated dataset used for training, validation, testing, and hyperparameters tuning contains a total of $477,\!141$ samples. Each sample consists of four traces, with $5,\!000$ values at respective time points. For each sample, one has the simulation-based (true) information available on several quantities, of which \logE{} and \Xmax{} are of primary physics interest. Table~\ref{tab:stats} shows the basic dataset statistics for these two variables.

\begin{table}[h]
\centering
\begin{tabular}{l|r|r}
\toprule
Sample size & \multicolumn{2}{c}{$477,\!141$} \\ \hline
Variable    & $\logE{}$ & \Xmax{} [$\unit{g/cm^2}$] \\ \hline
Mean & $19.50$ & $844.4$ \\
Standard deviation & $0.657$ & $63.81$ \\ \hline
Minimum & $17.50$ & $601.8$ \\
$25\%$ quantile & $19.17$ & $801.7$ \\
$50\%$ quantile & $19.69$ & $843.4$ \\
$75\%$ quantile& $20.11$ & $887.6$ \\
Maximum & $20.50$ & $1,\!000$ \\
\bottomrule
\end{tabular}
\caption{Selected statistical properties of the simulated EAS data sample used in the study for the two physics variables of the shower energy and the maximum shower development.}
\label{tab:stats}
\end{table}
 
It is evident that the \logE{} and \Xmax{} quantities have largely different scales, which can negatively impact the training process due to disparities in the loss function's values. 
In addition, the traces dataset is not in a suitable form, either: the trace signal values range from $0$ to $2,\!000$ with many zero bin contents which, in turn, may lead to issues during the training process of certain methods, particularly neural networks (NNs); see~\cite{Schmidhuber2022} for a historical review on NNs. 
This means that some form of preprocessing is needed, which is described in the next section.

\subsection{Data preprocessing}
Since the machine learning models discussed in the next section are based on NNs, with the exception of the benchmark method, the input and output data are normalized to the interval $[0,1]$. 
The dataset is first divided into training, validation, and testing subsets. 
Normalization parameters are determined using only the training and validation subsets and are then applied consistently to all subsets.
The quantities \logE{} and \Xmax{} are scaled using standard min-max normalization.
For the traces, normalization is performed using a single global maximum computed over all time bins, traces, and events in the training and validation samples.

\subsection{Machine learning tools and their architecture}
\label{subsec:models}
In this section, we introduce and test several models for predicting the values of \logE{} and \Xmax{}. The models were implemented and trained using the scikit-learn~\cite{Pedregosa2011}, Tensorflow~\cite{Abadi2015}, and Pytorch~\cite{Paszke2019} modules within Python3~\cite{Rossum2009}.

The first model (Model~I) is a Gradient Boosting Regression (GBR) model~\cite{Friedman20011189} and serves as a benchmark. We chose it because it ranks among the best-performing models that do not rely on NNs.

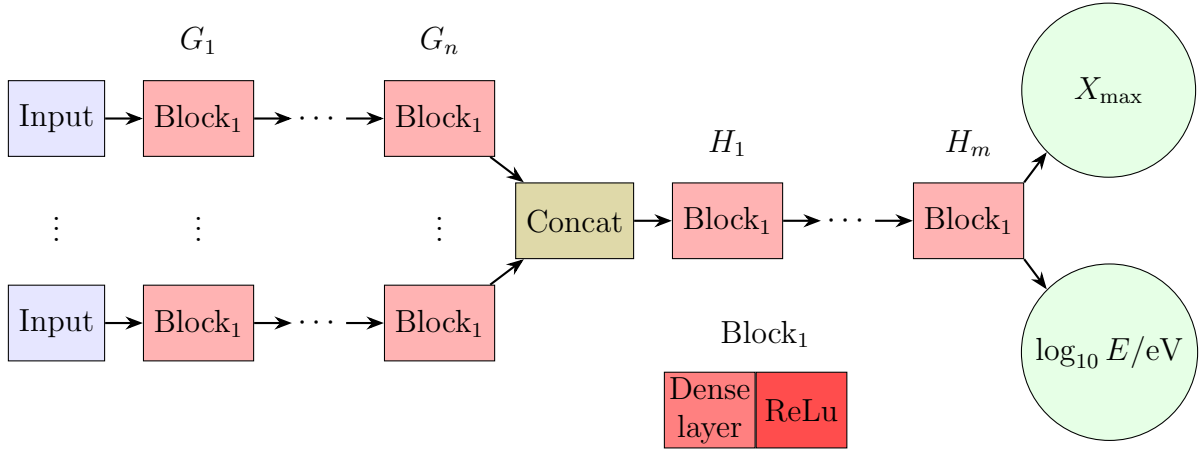
\begin{figure}[htb!]
\centering
\begin{tikzpicture}

\node[inputbox] (in1) {\inpt{1}};
\node[dotsnode, below=0.5cm of in1] (dots) {$\vcenter{\vdots}$};
\node[inputbox, below=0.5cm of dots] (in3) {\inpt{4}};

\node[block, right=0.5cm of in1] (hl1) {\blck{1}};
\node[block, right=0.5cm of in3] (hl3) {\blck{1}};
\node[above=0.2cm of hl1] {$G_1$};
\node[dotsnode, below=0.5cm of hl1] (dots) {$\vcenter{\vdots}$};

\draw[arrow] (in1) -- (hl1);
\draw[arrow] (in3) -- (hl3);

\node[dotsnode, right=0.5cm of hl1] (dots1) {$\cdots$};
\node[dotsnode, right=0.5cm of hl3] (dots3) {$\cdots$};

\draw[arrow] (hl1) -- (dots1);
\draw[arrow] (hl3) -- (dots3);

\node[block, right=0.5cm of dots1] (hln1) {\blck{1}};
\node[block, right=0.5cm of dots3] (hln3) {\blck{1}};
\node[above=0.2cm of hln1] {$G_n$};
\node[dotsnode, below=0.5cm of hln1] (dots) {$\vcenter{\vdots}$};

\draw[arrow] (dots1) -- (hln1);
\draw[arrow] (dots3) -- (hln3);

\path (hln1.north) -- (hln3.south) coordinate[midway] (enc_mid);

\node[block, right=1cm of enc_mid,fill=olive!30] (concat) {Concat};

\draw[arrow] (hln1) -- (concat);
\draw[arrow] (hln3) -- (concat);

\node[block, right=0.5cm of concat] (lay5) {\blck{1}};
\node[above=0.2cm of lay5] {$H_{1}$};
\draw[arrow] (concat.east) -- (lay5.west);
\node[dotsnode,right=0.5 of lay5] (lay6) {$\cdots$};
\draw[arrow] (lay5.east) -- (lay6.west);
\node[block,right=0.5cm of lay6] (lay7) {\blck{1}};
\node[above=0.2cm of lay7] {$H_{m}$};
\draw[arrow] (lay6.east) -- (lay7.west);

\node[output, above right=0.4cm and 0.3cm of lay7,minimum width=2.3cm] (xmax) {\Xmax{}};
\draw[arrow] (lay7.north east) -- (xmax.south west);

\node[output, below right=0.4cm and 0.3cm of lay7,minimum width=2.3cm] (loge) {$\logE{}$};
\draw[arrow] (lay7.south east) -- (loge.north west);

\node[below left=1cm and 0cm of lay6] (Blockdesc) {\blck{1}};
\node[block,below=0.5cm of Blockdesc.west,fill=red!50, minimum width=1.2cm, inner sep=0pt] (left) {Dense\\layer};
\node[block,right=0cm of left.east,fill=red!70, minimum width=1.2cm, inner sep=0pt] (right) {ReLu};

\end{tikzpicture}
\caption{Architecture of the Model II and III (feedforward NN) used in the study. Each block $H_i$ and $G_j$ represents a dense NN layer followed by a ReLu activation function. This describes an architectural pattern that allows for two distinct modes.
The first mode corresponds to immediate concatenation of the inputs, which are then passed directly into the NN. Model II has thus no parallel blocks before the Concat (concatenate) block. The second mode involves parallel processing of the inputs, followed by concatenation of the results, and then further NN processing (Model~III).}
\label{fig_arch1}
\end{figure}

Models~II and~III are both based on NNs with classical dense (fully-connected) layers, but their architectures differ in how they handle inputs; see Figure~\ref{fig_arch1}. Model~II concatenates all four traces into a single one of $20,\!000$ data points, which is then passed through the entire network. 
In contrast, Model~III processes the four individual traces (each with $5,\!000$ data points) in parallel. After a certain number of such separate layers, the outputs are concatenated and further processed by a standard NN. The idea behind this architecture is to try to separately capture some of the unique properties of each individual trace. The ReLu activation function~\cite{DBLP:journals/corr/abs-1803-08375} is utilised to introduce the nonlinearity required for solving the problem. 
Consequently, Model~III consists of both separate and common blocks of layers, while Model~II employs only common layers.

\begin{figure}[htb!]
\centering
\begin{tikzpicture}
\node[inputbox] (in1) {\inpt{1}};

\node[blockd, right=0.5cm of in1] (hl1) {\blck{2}};
\node[above=0.2cm of hl1] {$G_1$};

\draw[arrow] (in1) -- (hl1);

\node[dotsnode, right=0.5cm of hl1] (dots1) {$\cdots$};

\draw[arrow] (hl1) -- (dots1);

\node[blockd, right=0.5cm of dots1] (hln1) {\blck{2}};
\node[above=0.2cm of hln1] {$G_n$};

\draw[arrow] (dots1) -- (hln1);

\node[block, right=0.5cm of hln1,fill=brown!70] (concat) {Flatten};

\draw[arrow] (hln1) -- (concat);

\node[block, right=0.5cm of concat] (lay5) {\blck{1}};
\node[above=0.2cm of lay5] {$H_{1}$};
\draw[arrow] (concat.east) -- (lay5.west);
\node[dotsnode,right=0.5cm of lay5] (lay6) {$\cdots$};
\draw[arrow] (lay5.east) -- (lay6.west);
\node[block,right=0.5cm of lay6] (lay7) {\blck{1}};
\node[above=0.2cm of lay7] {$H_{m}$};
\draw[arrow] (lay6.east) -- (lay7.west);

\node[output, above right=0.4cm and 0.3cm of lay7,minimum width=2.3cm] (xmax) {\Xmax{}};
\draw[arrow] (lay7.north east) -- (xmax.south west);

\node[output, below right=0.4cm and 0.3cm of lay7,minimum width=2.3cm] (loge) {\logE{}};
\draw[arrow] (lay7.south east) -- (loge.north west);

\node[below=0.5cm of dots1] (Blockdescdva) {\blck{2}};
\node[block,below=0.5cm  of Blockdescdva.west,fill=orange!50,minimum width=1.2cm, inner sep=0pt] (left) {Conv\\ 1D};
\node[block,right=0cm of left.east,fill=orange!70,minimum width=1.4cm, inner sep=0pt] (right) {Avg\\ pooling};

\node[below=0.5cm of lay6] (Blockdesc) {\blck{1}};
\node[block,below=0.5cm of Blockdesc.west,fill=red!50,minimum width=1.2cm, inner sep=0pt] (left) {Dense\\layer};
\node[block,right=0cm of left.east,fill=red!70,minimum width=1.2cm, inner sep=0pt] (right) {ReLu};
\end{tikzpicture}
\caption{The architecture of Models IV and V consists of two distinct stages. In the first stage, each trace sample is represented as a matrix of the size $1\times 20,\!000$ or $4\times 5,\!000$ for Model IV and V, respectively, further processed by convolutional layers followed by average pooling. The resulting output is then flattened and passed to a standard NN with dense blocks for further processing. The architecture is flexible, designed to work with either four individual traces, each representing a distinct channel, or with their concatenation, which forms a single trace with one channel.}
\label{fig_arch2}
\end{figure}
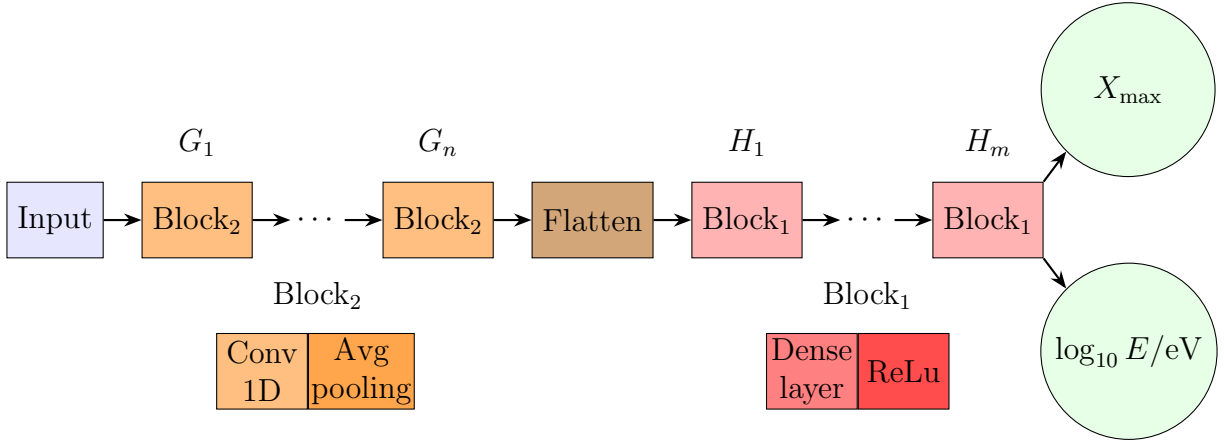

Models~IV and~V (see Figure~\ref{fig_arch2}) mirror the architectures of Models~II and~III but are designed to exploit the fact that traces can be treated as one-dimensional images. This means we treat each sample as either a $1\times 20,\!000$ image with a single channel (Model~IV), or as a $4\times5,\!000$ image (Model~V), \emph{i.e.} as an image with four channels. This representation allows to apply tools typically used for images, specifically the convolution and pooling, with a specified number of kernels and strides. We use a certain number of convolutional layers followed by four standard dense NN layers.

All models produce the same two outputs, specifically $\Xmax{}$ and~$\logE{}$. 
The input representation differs between the models: for Models~II and~IV, the input has the shape of $1\times~\!20,\!000$, while for Models~III and~V, it has the shape of $4\times 5,\!000$.

\subsection{Hyperparameters optimisation}
To determine the optimal hyperparameters configuration, a grid search was performed for the models introduced in Section \ref{subsec:models}.
To reduce computational and memory costs, the search was carried out using only half of the randomly selected simulation dataset.
This subset was then randomly divided into three disjoint parts -- training, validation, and testing -- with proportions of $40\%$, $10\%$, and $50\%$, respectively.
The same data split was used for all models.

Even with only the half of the full dataset, the average training time for a single hyperparameters configuration exceeded 
three hours for Models~II--V when using $4$ CPUs (the total number of configurations is $248$; see below).
In addition, the computational cost was significantly higher for Model~I, which required more than four times longer training on average than the best-performing Models~II--V identified during the hyperparameters optimisation.
Consequently, Model~I was not further optimised, as it was extremely time-consuming to train while providing no significant performance improvement.
Its results were approximately a factor of two worse than the median performance achieved by the other models during the grid search.

The search for suitable hyperparameters configurations for Models~II–V was performed over the parameter ranges listed in Table \ref{tab:hyperparameters}.
The optimal hyperparameters for each model were determined by comparing performance on the testing subset, which was not used during training nor validation.
The hyperparameter search was extended until the selected configuration was not located at the boundary of the explored parameter space, reducing the possibility that the optimum lay outside the tested ranges.

For Models~II and~III, the number of common layers (comm lyrs) and the number of neurons per layer (nrns) were varied. 
In Model~III, the number of separate layers (sep lyrs) was additionally varied, with the constraint that the number of common and separate layers was kept equal during the optimisation.
For Models~IV and~V, the hyperparameter search covered the number of convolutional layers (conv lyrs), the convolution kernel size (kern), and the number of neurons in the dense layers (nrns).
For all configurations, the number of dense layers was fixed to four.

\begin{table}[!ht]
    \centering
\begin{tabular}{l|cccc}
\toprule
 & sep lyrs & comm lyrs & nrns &    \\ \hline
Model II  & 0 & [2, 8] & [25, 150] & \\
Model III & [1, 6] & [1, 6] & [25, 150] & \\
\hline
\hline
 & conv lyrs & kern & dense lyrs & nrns \\ \hline
Model IV & [1, 8] & [3, 7] & 4 & [25, 500]  \\
Model V  & [1, 6] & [3, 7] & 4 & [25, 500]  \\
\bottomrule
\end{tabular}
\caption{The parameter ranges for Models~II-V. 
}
\label{tab:hyperparameters}
\end{table}

In order to evaluate the performance of individual configurations, we calculate the root mean squared error (RMSE) for \logE{} and \Xmax{} as
\begin{equation}
    \label{eq:RMSE}
    \mathrm{RMSE}_x =\sqrt{\frac{1}{N}\sum _{i=1}^N(x_i-\widetilde x_i)^2}\,,
\end{equation}
where $x$ denotes either \logE{} or \Xmax{}, $x_i$ is the true (generated) value of $x$ for the $i$-th simulation after rescaling to a dimensionless variable, and $\widetilde x_i$ is the corresponding predicted value.
During the optimisation process, we use the mean squared error (MSE) as the loss function calculated as
\begin{equation}
\label{eq:MSE}
\mathrm{MSE} = \mathrm{RMSE}_y^2 + \mathrm{RMSE}_z^2,
\end{equation}
where $y$ and $z$ represent \logE{} and \Xmax{}, respectively.

\begin{figure}[htb!]
    \centering
    \includegraphics[width=0.47\linewidth]{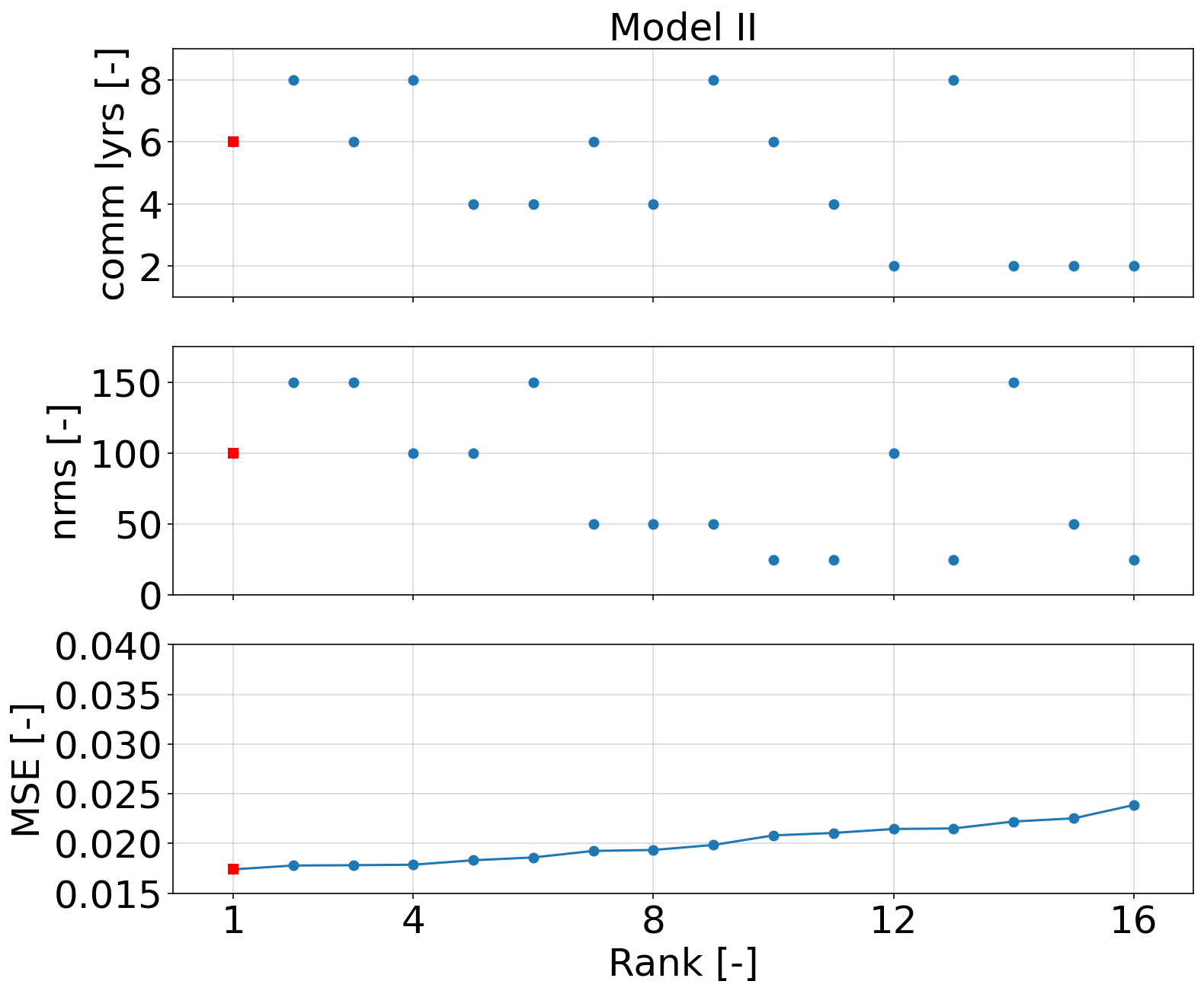}
    \includegraphics[width=0.47\linewidth]{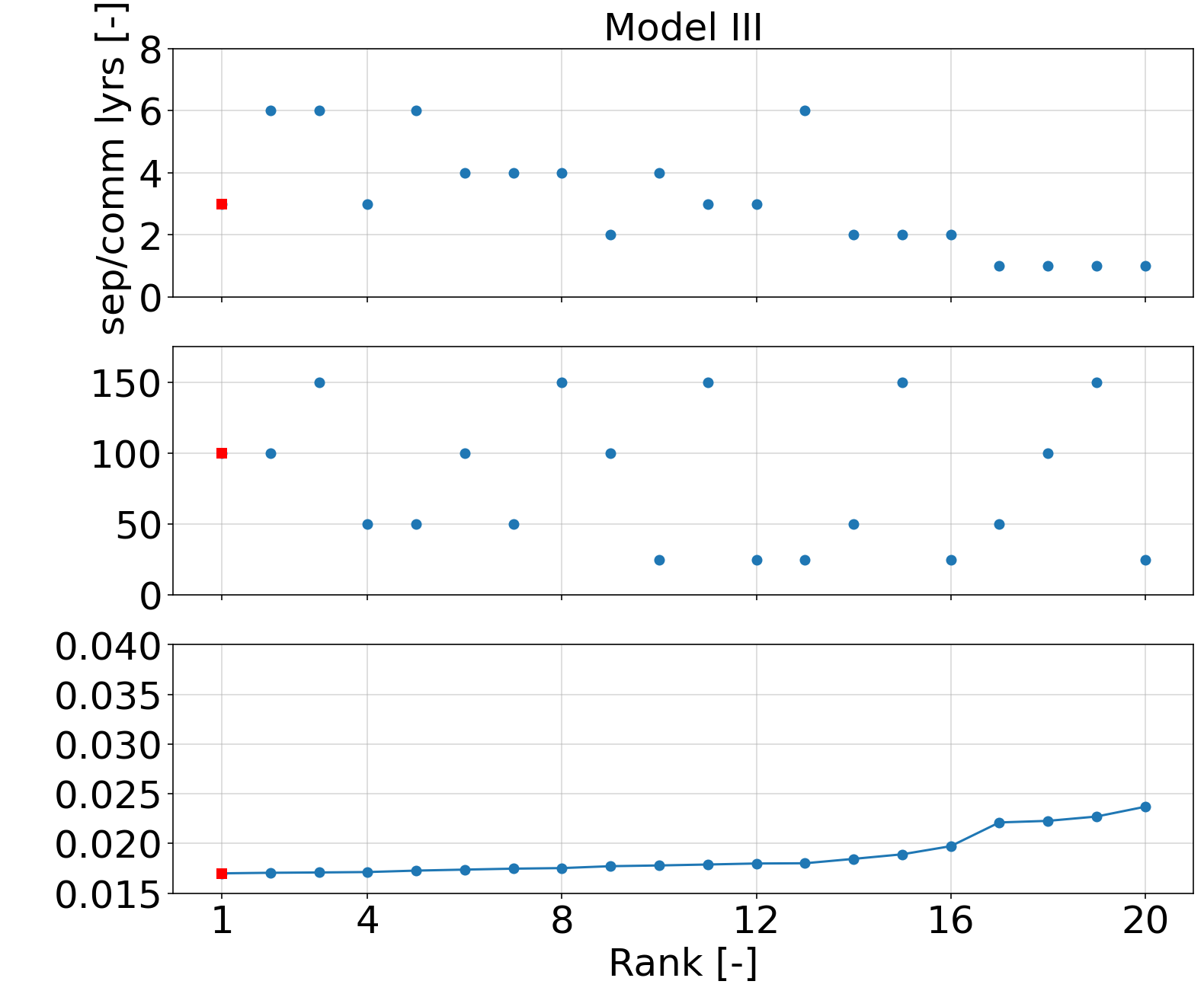}
    \caption{Ranking of tested hyperparameters configurations for Model~II (left) and Model~III (right) according to increasing MSE loss values, with the best-performing models corresponding to the lowest ranks. Optimal configurations are highlighted by red squares. See the text for details.
    }
    \label{fig:hyperoptimization_01}
\end{figure}

Figure \ref{fig:hyperoptimization_01} shows the ranking of all tested configurations for Model~II (left panel) and Model~III (right panel) according to the loss function MSE calculated on the testing subset.
For comparison, Model~I achieved an MSE of approximately $0.04$, indicating substantially worse performance than Models~II and~III.

For Model~II, the most suitable configuration (highlighted by a red square) consists of six common layers with $100$ neurons per layer.
Configurations with fewer common layers or fewer neurons resulted in worse performance.
It is also demonstrated that the selected optimal configuration is not located at the boundary of the tested parameter space, as similar performance was also achieved using eight common layers with $150$ neurons per layer. 

For Model~III, the optimal configuration corresponds to three separate and three common layers, each with $100$ neurons.
This configuration, highlighted by a red square in the right panel of Figure \ref{fig:hyperoptimization_01}, also does not lie at the boundary of the tested parameter space.

\begin{figure}[htb!]
    \centering
    \includegraphics[width=0.47\linewidth]{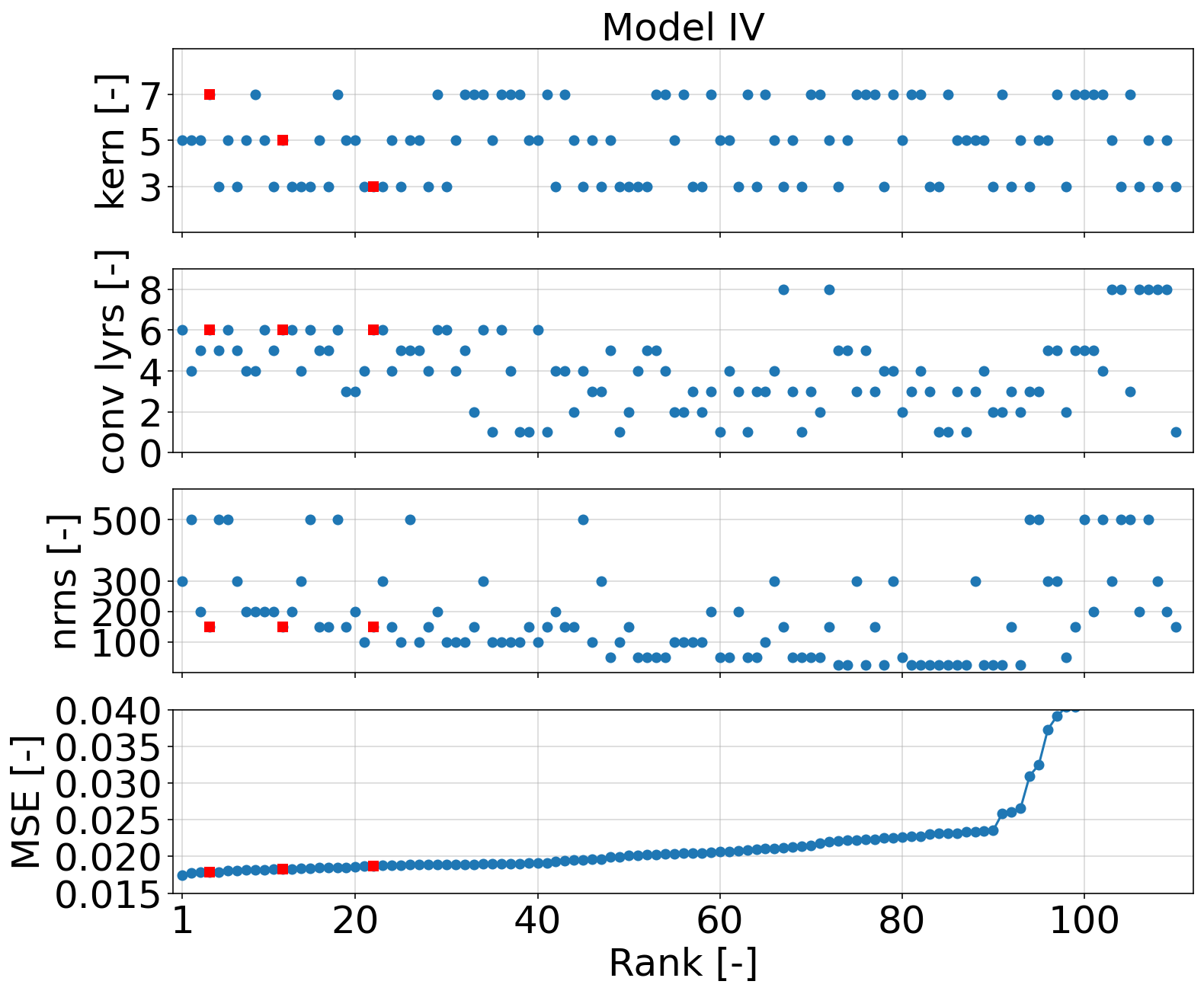}
    \includegraphics[width=0.47\linewidth]{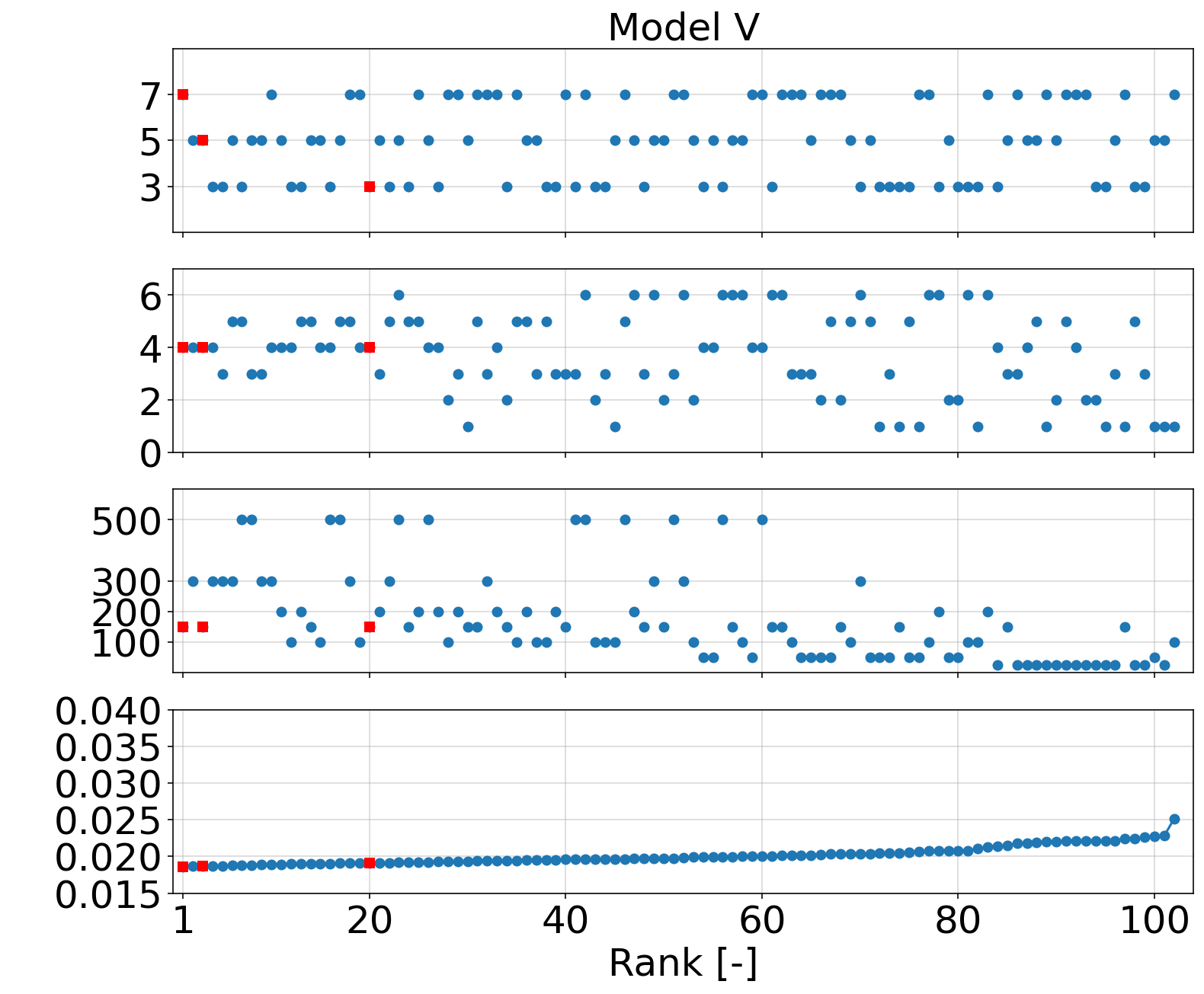}
    \caption{Ranking of tested hyperparameters configurations for Model~IV (left) and Model~V (right) according to increasing MSE loss values, with the best-performing models corresponding to the lowest ranks. Optimal configurations are highlighted by red squares. See the text for details.
    }
    \label{fig:hyperoptimization_02}
\end{figure}

Figure \ref{fig:hyperoptimization_02} shows the ranking of all tested configurations for Model~IV (left panel) and Model~V (right panel) according to the MSE.
The kernel size is found to have no impact on performance for both models. 
In contrast, the number of convolutional layers and the number of neurons per dense layer have a significant effect on performance.

One the one hand, configurations with too few convolutional layers or too few neurons per dense layer result in insufficient model capacity, making it impossible for the convolutional neural network to accurately predict energy and \Xmax{}.
On the other hand, very large numbers of convolutional layers and neurons substantially increase the number of trainable parameters, which complicates the optimisation process during training. 
This effect is evident for configurations with $300$ and $500$ neurons per layer in Model~IV, for which the resulting MSE varies significantly, depending on whether the optimisation converges to a suitable minimum in the high-dimensional parameter space. 
For clarity, the upper limit of the MSE loss values on the $y-$axis is set to $0.04$, as larger values are not informative any more. 

For Model~IV, the optimal number of convolutional layers is six. 
The best-performing configurations in terms of the number of neurons per dense layer lie in the range from $150$ to $500$. 
For the reasons discussed above, $150$ neurons were selected.
Since the kernel size is not a decisive factor, all three tested values (kern $= 3, 5, 7$) were considered.
These configurations were then trained and evaluated using the full simulation dataset, and the best-performing configuration was selected; this corresponds to a kernel size of three. 
However, all three configurations yield comparable performance, hence kernel sizes of five or seven could also be used without a significant loss in performance.

For Model~V, the optimal number of convolutional layers is four, with $150$ neurons per a~dense layer. 
The choice of the number of neurons follows the same reasoning as for Model~IV.
As in Model~IV, three configurations with kernel sizes of $3, 5$, and $7$ were evaluated using the full simulation dataset.
The best performance was obtained for a kernel size of five.

For clarity, the optimal hyperparameters configurations for Models~II–V are summarised in Table~\ref{tab:optimal}.
\begin{table}[!ht]
    \centering
\begin{tabular}{l|cccc}
\toprule
 & sep lyrs & comm lyrs & nrns &    \\ \hline
Model II  & 0 & 6 & 100 & \\
Model III & 3 & 3 & 100 & \\
\hline
\hline
 & conv lyrs & kern & dense lyrs & nrns \\ \hline
Model IV & 6 & 3 & 4 & 150  \\
Model V  & 4 & 5 & 4 & 150  \\
\bottomrule
\end{tabular}
\caption{The optimal hyperparameters configurations for Models~II–V.
Models~II and~III are optimised with respect to the number of separate and common layers, and number of neurons in each layer.
Models~IV and~V are optimised with respect to the number of convolution layers, the convolution kernel size, and the number of neurons in each dense layer.}
\label{tab:optimal}
\end{table}

\section{Results}
In this section, we introduce the results for all models using their optimal hyperparameters configurations given in Table \ref{tab:optimal}. 
The full simulation dataset is employed, using the same data split as in the hyperparameters optimisation, \textit{i.e.} the data are randomly divided into training, validation, and testing subsets with proportions of $40\%$, $10\%$, and $50\%$, respectively.
The results are presented for \logE{} and \Xmax{} separately.

To gain more insight into how well the prediction corresponds to the true values, we compute the Pearson correlation coefficient as
\begin{equation}
\rho_{x\widetilde x}=\frac{\sum _{i=1}^N(x_i-\bar x)(\widetilde x_i-\overline{\widetilde x})}{
\sqrt{\sum _{i=1}^N(x_i-\bar x)^2\sum _{i=1}^N(\widetilde x_i-\overline{\widetilde x})^2}}\,,
\end{equation}
where $N$ is the number of events in the set under study and $\bar x$ and $\overline{\widetilde x}$ denote the mean values of the true and predicted quantities, respectively, for either \logE{} or \Xmax{}.

\begin{figure}[htb!]
    \centering
    \includegraphics[width=0.82\linewidth]{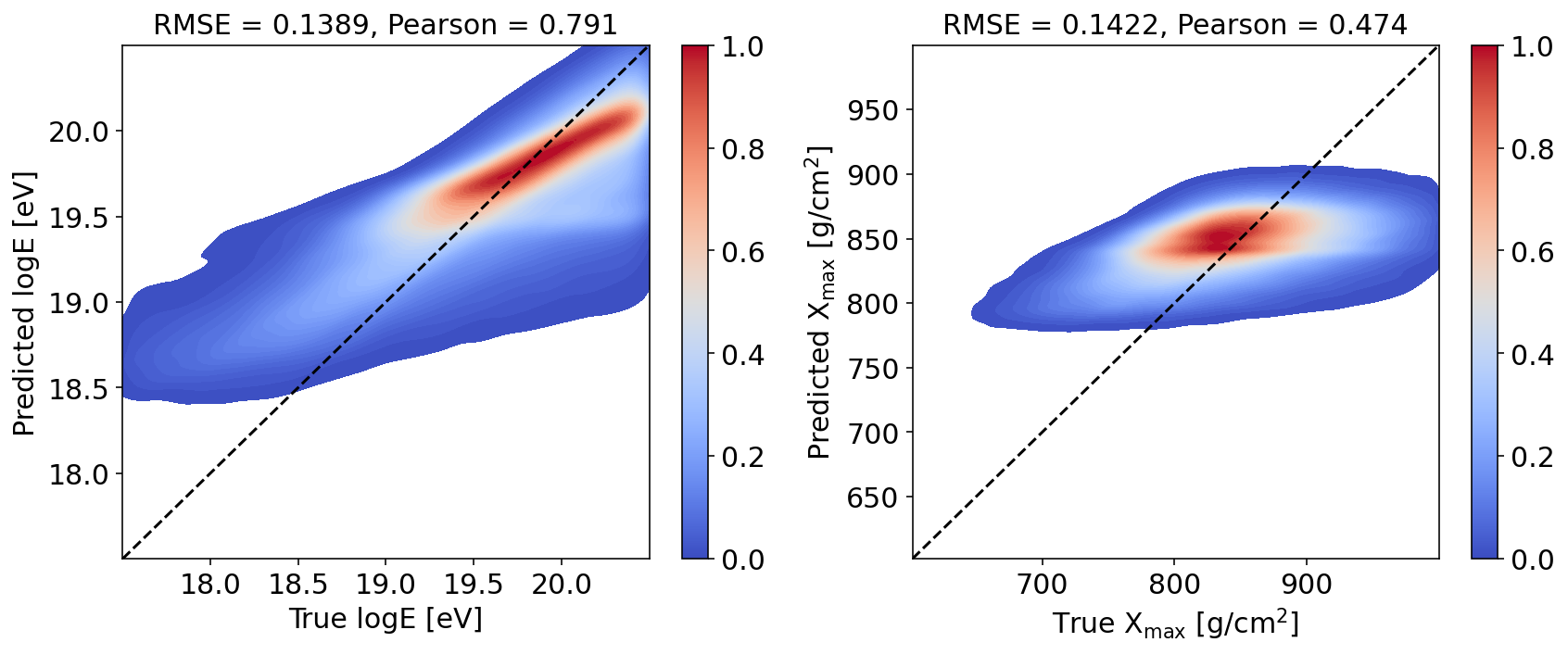}
    \caption{Results for the GBR classifier (Model~I) in terms of the correlation between the predicted ($y$-axis) and true ($x$-axis) values for the \logE{} (left) and the \Xmax{} (right) shower parameters. The figure also shows the RMSE (under the min-max scaling) and Pearson correlation coefficient calculated on the testing set, along with a dashed line representing the ideal performance. The relative density is indicated by the colour scale on the right.}
    \label{fig:results_GBR_I}
\end{figure}

The results for the benchmark (not optimised) GBR model are shown in Figure~\ref{fig:results_GBR_I}, exhibiting only a limited correlation between the predicted and true values of both physics variables. 
%The full simulation dataset was divided into training, validation, and testing subsets using the same proportions as for Models~II--V.

\begin{figure}[htb!]
    \centering
    \includegraphics[width=0.82\linewidth]{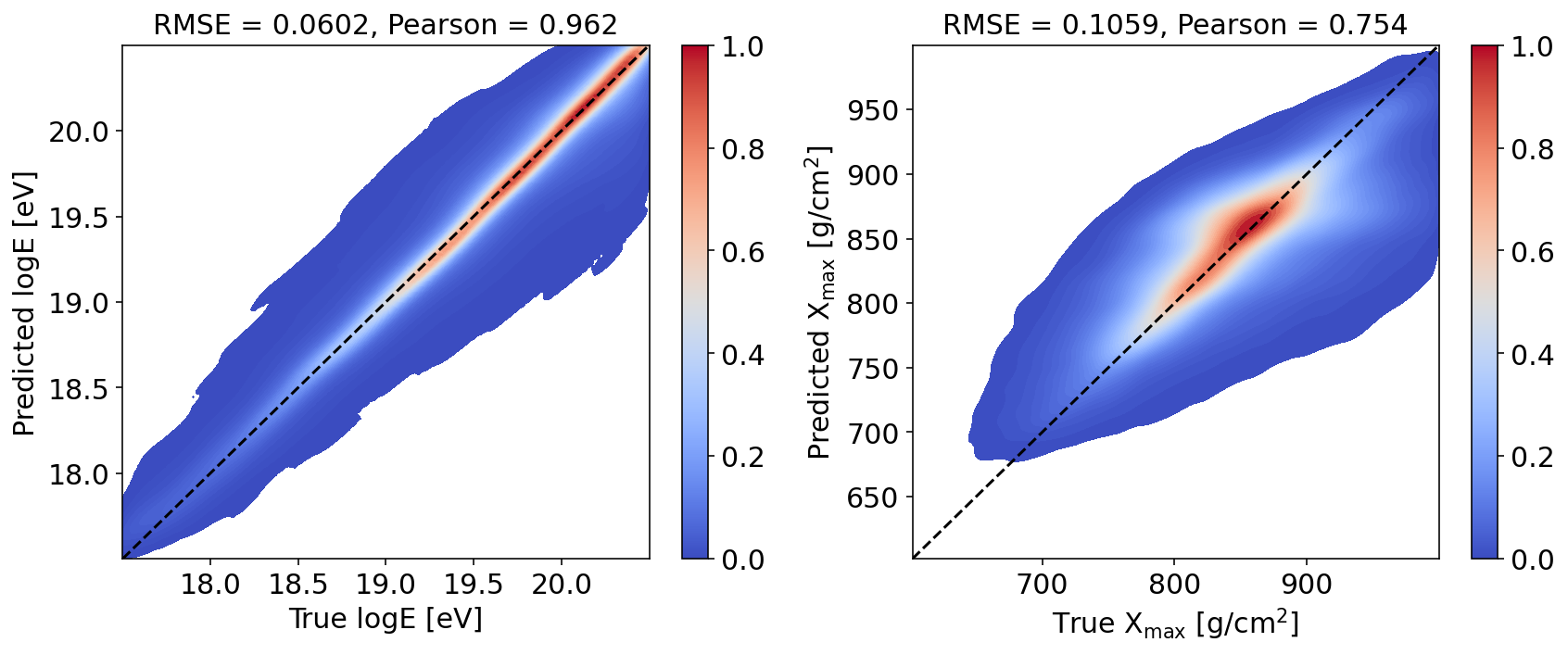}
    \caption{
    Results for Model II (NN using 6 common layers with 100 neurons in each layer) in terms of the correlation between the predicted ($y$-axis) and true ($x$-axis) values for the \logE{} (left) and the \Xmax{} (right) shower parameters. The figure also shows the RMSE (under the min-max scaling) and Pearson correlation coefficient calculated on the testing set, along with a dashed line representing the ideal performance. The relative density is indicated by the colour scale on the right.}
    \label{fig:results_NN_all_data_II}
\end{figure}

\begin{figure}[htb!]
    \centering
    \includegraphics[width=0.82\linewidth]{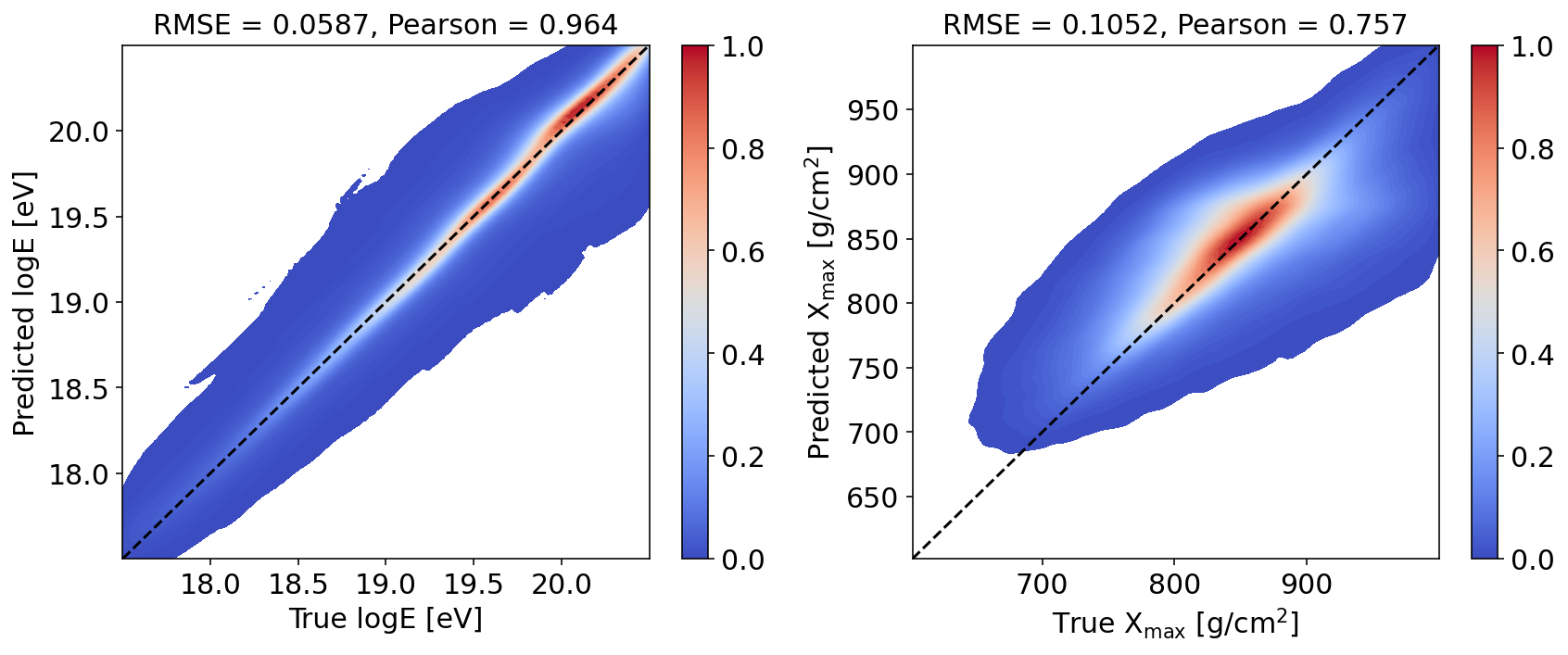}
    \caption{Results for Model III (NN with 3 separated layers, 3 common layers, and 100 neurons in each layer) in terms of the correlation between the predicted ($y$-axis) and true ($x$-axis) values for the \logE{} (left) and the \Xmax{} (right) shower parameters. The figure also shows the RMSE (under the min-max scaling) and Pearson correlation coefficient calculated on the testing set, along with a dashed line representing the ideal performance. The relative density is indicated by the colour scale on the right.} 
    \label{fig:results_NN_III}
\end{figure}

\begin{figure}[htb!]
    \centering
    \includegraphics[width=0.82\linewidth]{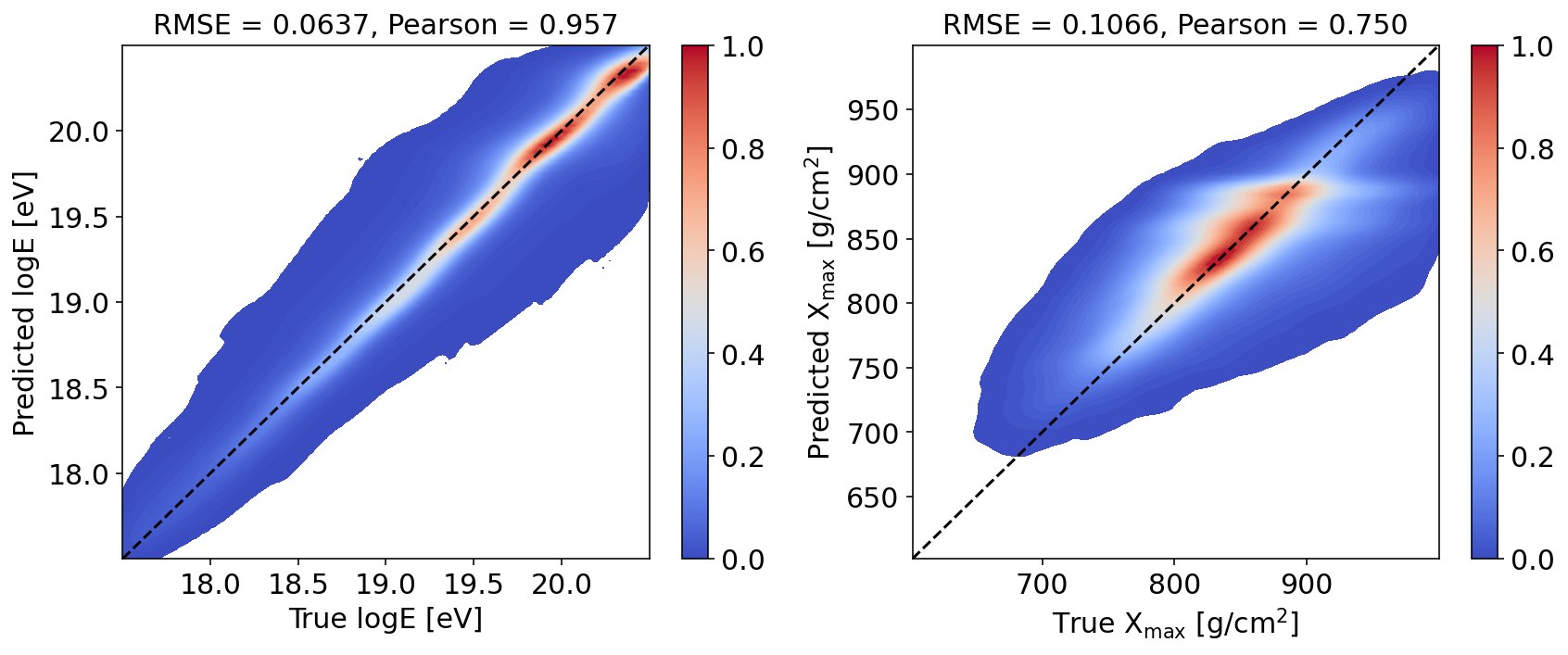}
    \caption{Results for Model IV (CNN using a single $1\times 20,\!000$ input processed by 6 convolutional layers and 4 dense NN layers, with 150 neurons in each layer) in terms of the correlation between the predicted ($y$-axis) and true ($x$-axis) values for the \logE{} (left) and the \Xmax{} (right) shower parameters. The figure also shows the RMSE (under the min-max scaling) and Pearson correlation coefficient calculated on the testing set, along with a dashed line representing the ideal performance. The relative density is indicated by the colour scale on the right.}
    \label{fig:results_CNN_IV}
\end{figure}

\begin{figure}[htb!]
    \centering
    \includegraphics[width=0.82\linewidth]{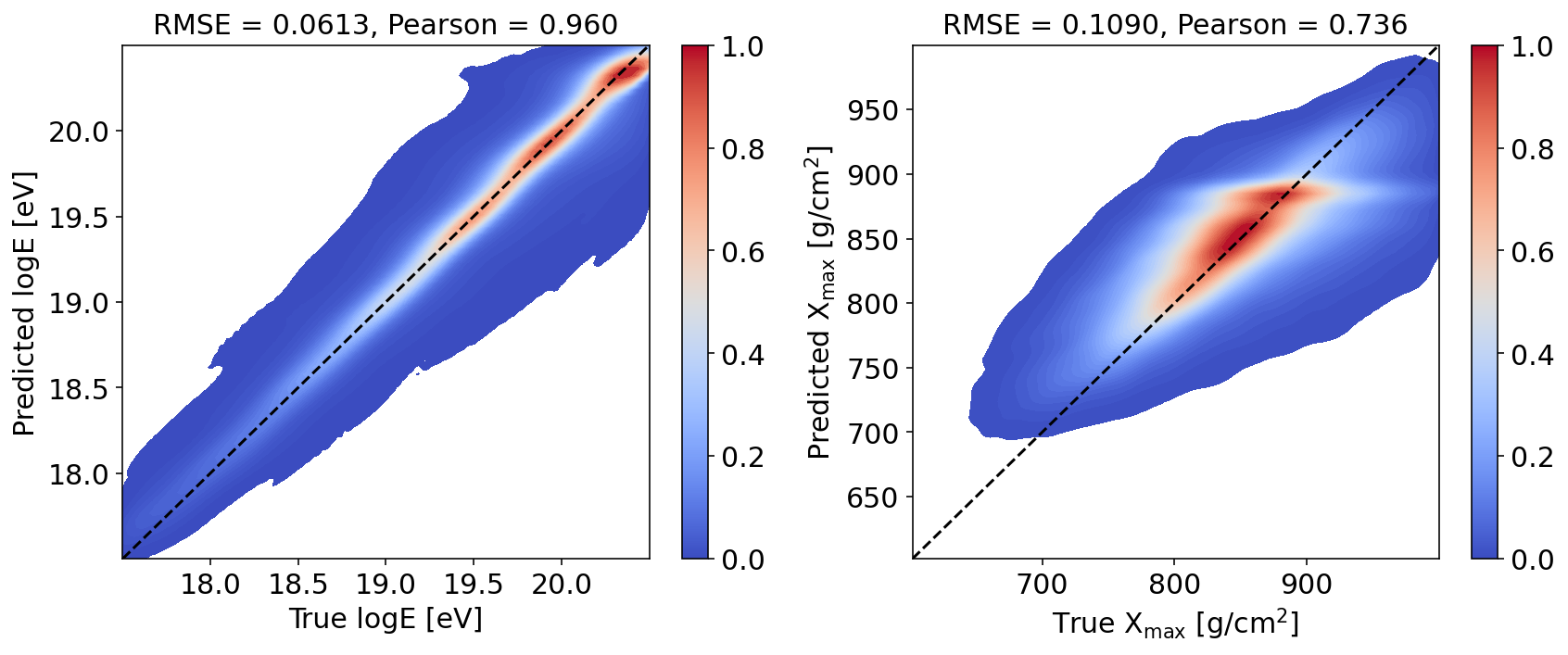}
    \caption{Results for Model V (CNN using $4\times 5,\!000$ input processed by 4 convolutional layers and 4 dense NN layers, with 150 neurons in each layer) in terms of the correlation between the predicted ($y$-axis) and true ($x$-axis) values for the \logE{} (left) and the \Xmax{} (right) shower parameters. The figure also shows the RMSE (under the min-max scaling) and Pearson correlation coefficient calculated on the testing set, along with a dashed line representing the ideal performance. The relative density is indicated by the colour scale on the right.}
    \label{fig:results_CNN_V}
\end{figure}

Next, results for all the NN classifiers are shown in Figures~\ref{fig:results_NN_all_data_II} -- \ref{fig:results_CNN_V}.
Specifically, Model~II is shown in Figure~\ref{fig:results_NN_all_data_II}, Model~III in Figure~\ref{fig:results_NN_III}, Model~IV in Figure~\ref{fig:results_CNN_IV}, and finally Model~V in Figure~\ref{fig:results_CNN_V}.

%\newpage
%~

%\newpage
The best networks resulted in a $96\%$ correlation between the predicted and true values of the energy in terms of \logE{}. The performance on the \Xmax{} variable is weaker, with a correlation of about~$75\%$.
The best performing models are compared in Table~\ref{tab:models_cmp} in terms of the Pearson correlation coefficients between the truth and the prediction, as well as in the precision in terms of the RMSE defined in Eq.~(\ref{eq:RMSE}) and the loss function MSE defined in Eq.~(\ref{eq:MSE}).

\begin{table}[!ht]
    \centering
\begin{tabular}{l|c|c|c|c|c}
\toprule
          & \multicolumn{2}{c|}{\logE{}} & \multicolumn{2}{c|}{\Xmax{}} & \multirow{2}{*}{MSE} \\ \cline{2-5}
          & RMSE  & $\rho$ & RMSE & $\rho$ & \\ \hline
Model I   & 0.1389 & 0.791  & 0.1422              & 0.474  & 0.0395 \\ 
Model II  & 0.0602 & 0.962  & 0.1059              & 0.754  & 0.0148 \\
Model III & 0.0587 & 0.964  & 0.1052              & 0.757  & 0.0145 \\
Model IV  & 0.0637 & 0.957  & 0.1066              & 0.750  & 0.0154 \\
Model V   & 0.0613 & 0.960  & 0.1090              & 0.736  & 0.0156 \\
\bottomrule
\end{tabular}
\caption{The Pearson correlation coefficients $\rho$, and RMSE and MSE (under the min-max scaling) for the best-performing models for the variables \logE{} and \Xmax{}.}
\label{tab:models_cmp}
\end{table}

All Models~II--V significantly outperform the benchmark Model~I in both \logE{} and \Xmax{}.
Although all these architectures perform similarly, CNNs are slightly outperformed by fully connected NNs.
Specifically, Model~III achieves approximately $6\%$ and $7\%$ higher performance in terms of MSE compared to Models~IV and~V, respectively.
CNNs are thus not necessary for this type of trace-based inference.
More importantly, artefacts in the form of weak periodic structures in the predictions of \Xmax{} and, to a smaller extend also of the \logE{} variable, are visible for the CNNs, which further emphasises the suitability of NNs over CNNs.

While Models~II and~III exhibit similar performance, Model~III is selected as the optimal model, as its performance in terms of MSE is approximately $2\%$ better than Model~II.
An additional reason for this choice is its more consistent behaviour during hyperparameters optimisation: among 20 best-performing configurations of Models~II and~III combined, 14 corresponds to Model~III.
However, Model~II could also be used if a different network architecture is preferred, as the performance difference is statistically insignificant.

\begin{figure}[htb!]
    \centering
    \includegraphics[width=0.80\linewidth]{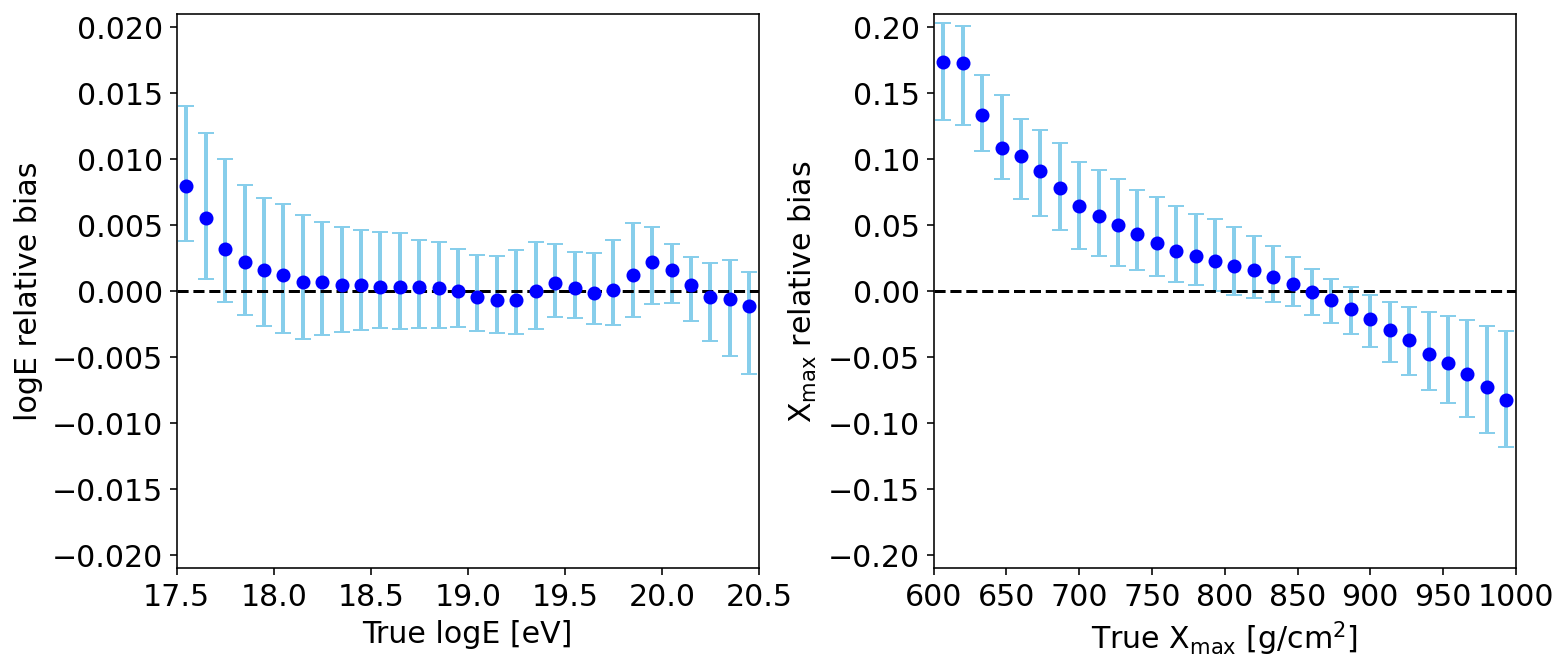} 
    \caption{The relative bias as a function of \logE{} (left) and \Xmax{} (right) variable for the best Model~III.
    The vertical bars in the bias plots indicate the $25$th and $75$th percentiles, while blue points indicate the median.}
    \label{fig:bias}
\end{figure}

We also present the analysis of the relative bias of the prediction compared to the true values of the shower parameters for the best Model~III.
We compute the relative bias of each variable as the predicted value minus the true value divided by the true value, \emph{i.e.} $\mathrm{(predicted - true)/ true}$. 
In Figure \ref{fig:bias}, the relative bias is shown as a function of true \logE{} and \Xmax{}. 
Blue points indicate the median, while the vertical bars indicate the $25$th and $75$th percentiles, \emph{i.e.} $50\%$ of data are in these intervals.  
The relative bias is below~$1\%$ for the \logE{} variable, while for the \Xmax{} it ranges from $+17\%$ down to $-8\%$.
For both parameters, predictions tend to slightly overestimate small true values and underestimate large true values. This can be attributed to the limited parameter space in the energy, and to the sparsely populated parameter space in the extreme tails of the \Xmax{}.

\begin{figure}[!ht]
    \centering
    \includegraphics[width=0.80\linewidth]{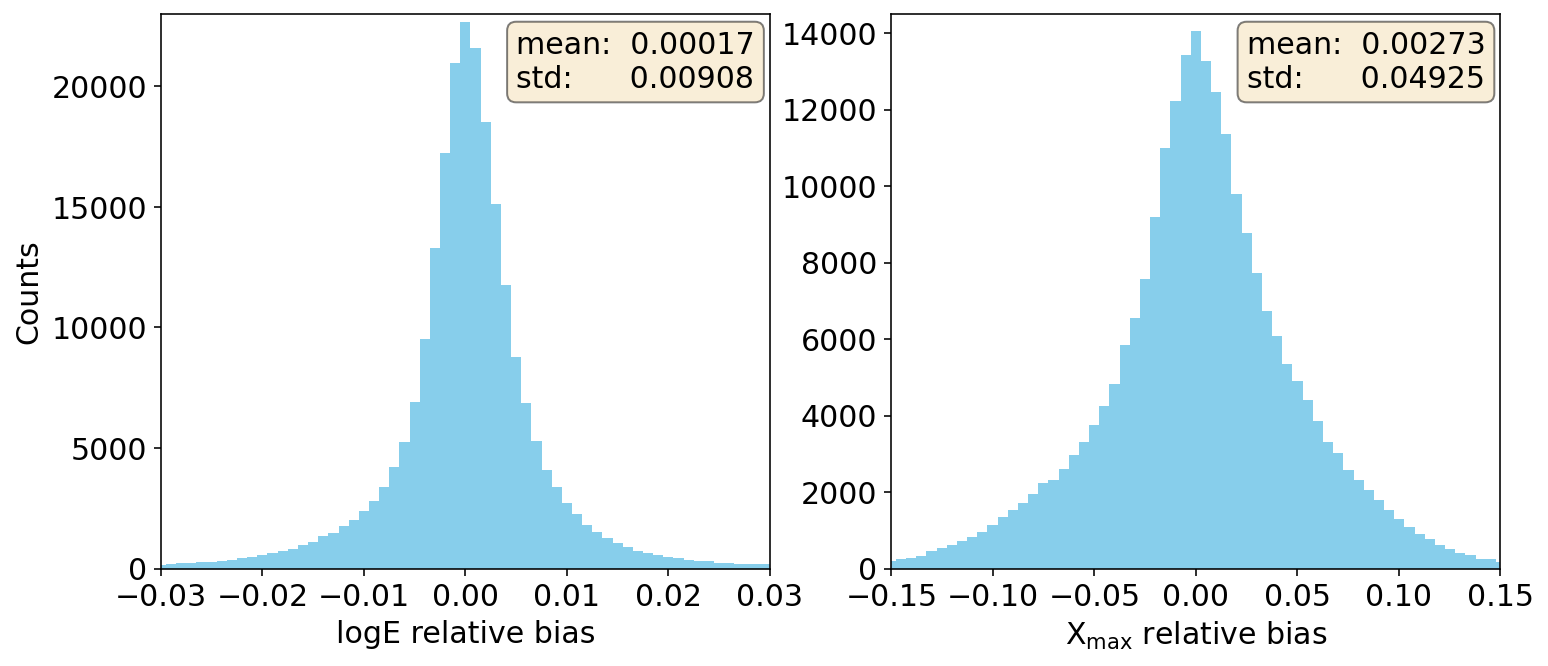} \\ 
    \caption{Histogram of the relative bias for the two shower parameters, \logE{} (left) and \Xmax{} (right) -- projected for all events in the testing sample (half of the total event set) for the best-performing Model~III.
    Indicated is also the mean and the standard deviation of the distribution.}
    \label{fig:bias_integrated}
\end{figure}

The distribution of the relative bias across all events is presented in Figure~\ref{fig:bias_integrated}, together with the corresponding mean and standard deviation.
While the mean is consistent with zero, signifying no overall relative bias, the standard deviation carries information on the relative resolution of the prediction, being $\approx 0.9\%$ for the \logE{} variable and $\approx 5\%$ for \Xmax{}.
The median of the absolute bias across all events, calculated as $\mathrm{abs(predicted - true)/ true}$, is $0.0031$ for \logE{}, while for \Xmax{} it is about eight times larger, equal to $0.0263$. 
However, even for \Xmax{}, the absolute bias remains below $3\%$ on average, demonstrating the good performance of the best Model III.

\newpage
\section{Discussion}
Table~\ref{tab:rhos_cmp} presents the Pearson correlation coefficients and MSE loss values obtained using either half or the full simulation dataset for training, validation, and testing.
Using the full dataset leads to a substantial reduction of the MSE for Models~II-V (by approximately $15\%$), as well as to stronger correlations between the predicted and true values of \logE{} and \Xmax{}.
This suggests that further performance improvements could be achieved with a larger simulation dataset; however, this possibility was not explored due to limitations in the available data. 
Still, the current results demonstrate the potential of the approach and already indicate sufficient performance.
In contrast, the benchmark Model~I exhibits similar performance regardless of the dataset size used, reflecting the model’s limited capability.

\begin{table}[!ht]
    \centering
\begin{tabular}{l|cc|cc|cc}
\toprule
 & \multicolumn{2}{c|}{\logE{}} & \multicolumn{2}{c|}{\Xmax{}} & \multirow{2}{*}{MSE$_\mathrm{half}$} & \multirow{2}{*}{MSE$_\mathrm{full}$} \\ \cline{2-5}
    & $\rho_\mathrm{half}$ & $\rho_\mathrm{full}$& $\rho_\mathrm{half}$ & $\rho_\mathrm{full}$  \\ \hline
Model I   & 0.788 & 0.791  & 0.475 & 0.474  & 0.0396 & 0.0395 \\
Model II  & 0.951 & 0.962  & 0.714 & 0.754  & 0.0174 & 0.0148 \\
Model III & 0.954 & 0.964  & 0.712 & 0.757  & 0.0170 & 0.0145 \\ 
Model IV  & 0.947 & 0.957  & 0.708 & 0.750  & 0.0187 & 0.0154 \\
Model V   & 0.951 & 0.960  & 0.698 & 0.736  & 0.0187 & 0.0156 \\
\bottomrule
\end{tabular}
\caption{
Pearson correlation coefficients $\rho$ and MSE loss values (under the min-max scaling) evaluated on the testing sets for Models~I--V.
The indices indicate whether half or the full simulation dataset was used for training, validation, and testing.
}
\label{tab:rhos_cmp}
\end{table}

The \Xmax{} parameter exhibits worse resolution in the extreme tails, as shown in Figure~\ref{fig:bias}, which can again be attributed to the sparsely populated parameter space in these regions.
While one might consider generating additional simulations for very low and high \Xmax{} values, this was not followed due to the fact that the lack of data in the tails reflects the realistic probability density functions of \Xmax{} (see Figure \ref{fig:conex_Xmax_profiles}).
This limitation is clearly a drawback of using a realistic distribution of \Xmax{}.
On the one hand, such distribution provides a physically motivated correlation between energy and \Xmax{} during training.
On the other hand, it results in fewer simulations at the extremes, leading to a lower resolution therein.
An alternative approach could be to test a uniformly distributed \Xmax{} and compare the resulting NN performance with Models~I--V; however, such a comparison is beyond the scope of this study.

The implications of the presented models for a single FAST telescope for the possible future larger arrays of FAST telescopes lay in the fact that the presented NNs can be used for cases where some parts of the array are not operational or prevented to observe the sky, \emph{e.g.} due to clouds or other atmospheric conditions. In these cases, the usage of the presented ML classifier architectures may retain the observatory capacity to measure the EAS energy spectrum although shower origin in terms of angles cannot be reliably inferred without the stereo geometry of more FAST telescopes. 

Next steps will include the implementation of realistic noise which can be obtained in situ for each of the observatories as it may depend on the location and time, having a temperature-dependent component as well as a component related to fluctuations in the night-sky background. One can apply the current algorithms to simulations overlaid by realistic noise, check their performance and then apply the technique to real data, both to raw signals or after some procedure of noise reduction, \emph{e.g.} similar to the noise treatment in the trigger design~\cite{KMEC2026110063}.
After such validation, the presented algorithms could be directly used for the actual data from the several FAST telescopes which are installed at the Auger and TA sites and are already taking data.

While simulations with realistic noise represent a natural next step, incorporating atmospheric non-homogeneities into NN training remains challenging, as the large number of possible atmospheric configurations cannot be adequately sampled with a reasonable number of simulations.
This limitation does not pose a problem when the NN is used to provide an initial estimate of EAS parameters while the final top-down reconstruction accounts for atmospheric non-homogeneities during the likelihood optimisation. 
However, it becomes a concern if the NN is intended to serve as a full reconstruction tool.
In such cases, for events affected by significant atmospheric non-homogeneities, the NN output should be used only as a first estimate of the EAS parameters.

One of the aims of this analysis was to identify an appropriate network architecture suitable for future extensions of the FAST telescope array.
Such developments may include training models for larger arrays of FAST telescopes operating in stereo mode, with the capability to predict a broader set of EAS parameters. 
The FAST mini-array deployment is planned for 2026, making these developments a key focus of possible future work.

\section{Conclusions}
Focusing on air showers with a realistic correlation between the shower energy and the maximum of the EAS development, we prove that the reconstruction of these two main physics-relevant parameters is possible using a single FAST telescope comprising just four photomultipliers. 
We obtain practically useful accuracy by making use of the time granularity and employing standard artificial NNs or their convolutional version. 

For the best model, an excellent correlation of $96\%$ and a~resolution below~$1\%$ is found for the reconstructed energy in terms of \logE{}.
The performance for \Xmax{} is slightly worse, with a correlation of~$76\%$ and a resolution of~$5\%$, corresponding to about~$40\,\unit{g/cm^2}$ for \Xmax{} of~$850\,\unit{g/cm^2}$.
These results are promising given that only a single FAST telescope is used, albeit without the noise. 
They provide a foundation for future studies incorporating realistic noise and multiple FAST telescopes,
and demonstrate that NNs can serve as effective tools for providing initial estimates in top-down reconstruction, or potentially, as its full replacement.

\section*{Acknowledgements}

The authors gratefully acknowledge the support of the Czech Science Foundation project GACR 23-07110S and the project of the Palacky~U no. IGA\_PrF\_2025\_015.

\noindent
We thank dr.~J.~Kmec and dr.~P.~Hamal for their insightful comments; we also thank prof. O.~Haderka and dr. D.~Mandát and members of the FAST collaboration for their useful remarks.

\section*{Keywords}
Astroparticle physics, Ultra-high-energy-cosmic rays, Machine learning, Supervised learning, Regression, Parameters inference, Fluorescence telescopes, Extensive Air Showers.

\section*{Code Availability}
The code corresponding to the traning as well as the trained models themselves can be accessed via the github respository~\cite{Tomecek_code}.

\bibliography{main}
\bibliographystyle{unsrt}

\end{document}